\documentclass[prl,aps]{revtex4}
\usepackage{epsfig}

\begin{document}

\title{Programmable quantum state discriminator by Nuclear Magnetic Resonance} 
\author{T. Gopinath, Ranabir Das, and Anil Kumar \\
{\small \it NMR Quantum Computing and Quantum Information Group.\\Department of Physics, and Sophisticated Instruments Facility.\\
Indian Institute of Science, Bangalore - 560012, India}}.

\begin{abstract}
In this paper a programmable quantum state discriminator is implemented by using nuclear magnetic resonance. We use a two qubit spin-1/2 
system, one for the data qubit and one for the ancilla (programme) qubit. This device does the unambiguous (error free) discrimination of 
pair of 
states of the data qubit that are symmetrically located about a fixed state. The device is used to discriminate both, linearly polarized
 states and elliptically polarized states. The maximum probability of the successful 
discrimination is achieved by suitably preparing the ancilla qubit. It is also shown that, the probability of discrimination depends on 
angle of unitary operator of the protocol and ellipticity of the data qubit state.
\end{abstract}

\maketitle

\section{I. Introduction}
 Researchers have studied the possibility of performing computations using quantum systems 
and conjectured 
that a machine based on quantum mechanical principles might be able to solve certain types of problems more efficiently than can be done 
on conventional computers\cite{fey,ben,deu}. Later Lloyd proposed that such a quantum computer might be built from an array of coupled two 
state 
quantum systems\cite{loy}.
 Its theoretical possibility has generated a lot of enthusiasm for its experimental realization\cite{pw,deujoz,gr,gru,db,ic}. In parallel with 
quantum computation, the related field of quantum information theory is developed, which forms the quantum analogue of classical information theory \cite{cov}. Several 
techniques are being exploited for quantum computing and quantum information processing, including nuclear magnetic resonance \cite{nmr1,nmr2,nmr3,nmr4,nmr5,nmr6,nmr7,nmr8}.    
 
   Recently quantum state discrimination has been studied extensively in the context of quantum communication and quantum cryptography 
\cite{phil,zh,barr,che,che1,hel,iva,peres}. Quantum state discrimination is the problem of determining the quntum state, given the 
constraint that it belongs to the previously specified set of non-orthogonal states. One of the characteristic features of 
quantum mechanics is that, it is impossible to devise a measurement that can distinguish non-orthogonal states perfectly \cite{ic}. 
However one can distinguish them with a finite probability by appropriate measurement strategy.
 There are two different optimal strategies of discrimination: (i) Probabilistic discrimination
(conclusive result, but error may appear) and (ii) Unambiguous discrimination(inconclusive result may appear, but no error). 
The unambiguous discrimination of two pure states was investigated by Ivanovic \cite{iva}, and the optimal procedure was given by 
Peres \cite{peres}.
Cheffels and Barnett have generalized Peres's solution to an arbitrary number of equally probable states which are related by a symmetry 
transformation\cite{che3}. The first experiment to discriminate two non-orthogonally polarized single photons states of light was done by Huttner et.al\cite{hut}.
 
   Quantum measurement is the final step of any quantum computation. 
In many situations the choice of an optimal measurement
depends on the task to be performed.  
A quantum multimeter is a quantum measurement device which can perform a specific class of generalized
measurements in such a way that each member of this class is selected by a particular quantum state of a programme register 
\cite{dus,fil,fiu,paz,ari}.
 The parameters determining the character of quantum measurement 
can be encoded in 
a quantum state of a programme register\cite{niel,vid,hil}. 
Dusek et. al \cite{dus} have shown that pair of non-orthogonal states of a qubit can be discriminated and the measurement to be done for this discrimination is decided by 
the state of the programme qubit (ancilla qubit). One can discriminate several pairs of states of a qubit by using the same protocol\cite{dus}. 
Such  
a quantum device is known as quantum multimeter for the discrimination of pair of qubit states. 
 Recently Dusek et. al \cite{dus1} have also demonstrated experimentally the possibility 
to control the discrimination process by the quantum state of ancilla
qubit, in linear optics by performing the partial measurement in the Bell basis. Cryptographic applications of quantum state discrimination  have been extensively studied in the literature\cite{dus2,ham,yon,zhu}.

    Nuclear magnetic resonance (NMR) has played a leading role for practical demonstration of quantum algorithms and gates
\cite{nmr1,nmr2,nmr3,nmr4,nmr5,nmr6,nmr7,nmr8}. . The unitary operators needed for implementation of these quantum circuits have mostly been realized 
using spin selective as well as transition selective radio frequency pulses and coupling evolution, utilizing spin-spin (J) or dipolar couplings among the 
spins \cite{nmr1,nmr2,nmr3,nmr4,nmr5,nmr6,nmr7,nmr8} . In this paper we demonstrate the implementation of quantum state discriminator which discriminates 
the pair of non orthogonal states as well as 
orthogonal states  which are symmetric about a particular state, conditioned on the state
of the ancilla qubit. We use spin selective pulses and evolution under J-coupling for the implementation.
Projective measurement required for the discrimination is simulated by a method given by Collins\cite{col}.
Our experimental results are in agreement with the theoretical results\cite{dus}.
To the best of our knowledge this is the first experimental demonstration of programmable quantum state discriminator by NMR.

        In section(II), we discuss the theory of discrimination of both elliptically and linearly polarized states. Experimental details
and results of different experiments are given in section(III). Results are concluded in section(IV). In the Appendix, unitary operators of ideal pulses are derived.  
\section{II. Theory}
The following protocol discriminates pair of elliptically polarized states of the data qubit unambiguously (error free).  
Let the two  states $\vert \psi_1 \rangle$ and $\vert \psi_2 \rangle$ of the data qubit  be (fig. 1a),
\begin{eqnarray}
\vert\psi_1\rangle=(x cos\theta_1+iy sin\theta_1)\vert0_D \rangle\ + (x sin\theta_1-iy cos\theta_1)\vert1_D \rangle, 
\nonumber \\
\vert\psi_2\rangle=(x cos\theta_1+iy sin\theta_1)\vert0_D \rangle\ - (x sin\theta_1-iy cos\theta_1)\vert1_D \rangle.
\end{eqnarray}
 Ellipticity $(\epsilon)$ of the data qubit states is defined as, tan$(\epsilon)$=y/x. Here y=x corresponds to circularly polarized states and 
y=0 ($\epsilon$= 0) corresponds to linearly polarized states (fig. 1b). The protocol uses one ancilla (programme) qubit for the discrimination. 
A quantum circuit for the discrimination is shown in Fig. 2. In this circuit the first qubit is the data qubit ($\vert \psi_D \rangle$) and the second qubit is 
the ancilla qubit ($\vert \psi_A \rangle$). The data qubit can be either $\vert \psi_1 \rangle$ or $\vert \psi_2 \rangle$. The aim of the protocol (fig. 2) is to determine whether the data qubit is
 $\vert \psi_1 \rangle$ or $\vert \psi_2 \rangle$, knowing the angle between $\vert \psi_1 \rangle$ and $\vert \psi_2 \rangle$. It is shown that, a pair of data qubit states $\vert \psi_1 \rangle$ and $\vert \psi_2 \rangle$ can be discriminated by suitably preparing the ancilla qubit. 
One can switch the apparatus
to work with several different pairs of data qubit states. In this paper we experimentally demonstrate the discrimination of both elliptically and
linearly polarized states, and compare the results with simulations. In the following, we first describe the discrimination of elliptically 
polarized states and later the linearly polarized states.

 Elliptically polarized states $\vert \psi_1 \rangle$ and $\vert \psi_2 \rangle$ (eqn. 1) can be re-written as,
\begin{eqnarray}
\vert \psi_1\rangle=a_1\vert 0 \rangle+b_1\vert 1 \rangle,
\nonumber \\
\vert \psi_2\rangle=a_1\vert 0 \rangle-b_1\vert 1 \rangle,
\end{eqnarray}
where $a_1= x cos\theta_1+iy sin\theta_1$ and $b_1=x sin\theta_1-iy cos\theta_1$ are complex numbers, where by definition $x^2+y^2=1$.

  $a_1$ and $b_1$ can also be written in polar form as, 
\begin{eqnarray}
a_1= e^{i \phi_1} cos (\eta)\hspace{0.5cm} and \hspace{0.5cm}b_1= e^{i \phi_2} sin (\eta),\hspace{0.3cm}where, 
\end{eqnarray}
 
\begin{eqnarray}
tan(\eta)= \frac{\vert b_1 \vert}{\vert a_1 \vert},
\hspace{0.2cm} & tan(\phi_1)&=(\frac{y sin\theta_1}{x cos\theta_1}),\hspace{0.2cm} and \hspace{0.2cm} tan(\phi_2) =(\frac{-y cos\theta_1}{x sin\theta_1}). 
\end{eqnarray}
   Then $\vert \psi_1 \rangle$ and $\vert \psi_2 \rangle$ can be written as general states on the Bloch sphere \cite {ic},
  
\begin{eqnarray}                                                                                                                            
\vert \psi_1\rangle=cos(\eta) \vert 0 \rangle +  e^{i\phi} sin(\eta)\vert 1 \rangle,
\nonumber \\
\vert \psi_2\rangle=cos(\eta) \vert 0 \rangle -  e^{i\phi} sin(\eta)\vert 1 \rangle,                                                 
\end{eqnarray}  
where $\phi= \phi_2 -\phi_1$ with the overall phase ($e^{i \phi_1}$) being neglected.

   Let the ancilla qubit (programme qubit) be,
\begin{eqnarray}
\vert \psi_A\rangle=a_2\vert 0 \rangle+b_2\vert 1 \rangle, 
\end{eqnarray}
where $a_2= x cos\theta_2+iy sin\theta_2$ and $b_2=x sin\theta_2-iy cos\theta_2$.
To discriminate $\vert \psi_1 \rangle$ and $\vert \psi_2\rangle$, the condition on $a_2$ and $b_2$ is derived as follows.

  The total input state is $\vert \psi_{DA}\rangle$ = $\vert \psi_D\rangle \otimes \vert \psi_A\rangle$, where the data qubit $\vert \psi_D \rangle$ is 
either $\vert \psi_1 \rangle$ or $\vert \psi_2 \rangle$ (eqn. 2).
 \begin{eqnarray}
\vert\psi_{DA}\rangle =(a_1a_2\vert 0_D0_A \rangle+a_1b_2\vert0_D1_A \rangle) \pm (b_1a_2\vert1_D0_A\rangle + b_1b_2\vert1_D1_A\rangle).
\end{eqnarray}
The sign of the second term in Eqn. (7) determines whether the data qubit is $\vert \psi_1 \rangle$ or $\vert \psi_2 \rangle$. The protocol for the discrimination requires  a unitary 
transformation on $\vert \psi_{DA}\rangle$ given by \cite{dus},

\begin{eqnarray}
U=\pmatrix{ cos(\alpha)&-sin(\alpha)&0&0 \cr sin(\alpha)&cos(\alpha)&0&0 \cr 0&0&1&0 \cr 0&0&0&1 },
\end{eqnarray}
\vspace{0.3cm}
where $\alpha$ is a fixed parameter which does not depend on the data and programme qubits states. The unitary operator given in Eqn. 8, is a rotation in the subspace spanned by $\vert 0_D 0_A \rangle$ and $\vert 0_D 1_A \rangle$, which
is achieved here by two controlled-not gates and four single qubit gates (fig. 2), where the
Single qubit gates are given by,

\vspace{0.3cm}
\begin{eqnarray}                                                                                                                            
X=\pmatrix{ 0&1 \cr 1&0 }, u1=\pmatrix{cos(\alpha/2)&sin(\alpha/2) \cr -sin(\alpha/2)&cos(\alpha/2)},  u2=\pmatrix{cos(\alpha/2)&-sin(\alpha/2) \cr 
sin(\alpha/2)&cos(\alpha/2)},
\end{eqnarray}
\vspace{0.3cm}
And controlled-not(CNOT) gate is given by,
\vspace{0.3cm}
\begin{eqnarray}
CNOT=\pmatrix{1&0&0&0 \cr 0&1&0&0 \cr 0&0&0&1 \cr 0&0&1&0}. 
\end{eqnarray}
\vspace{0.3cm}
After the application of the unitary transformation U(eqn. 8), the final state is,
\begin{eqnarray}
U\vert\psi_{DA}\rangle=(a_1a_2 cos\alpha - a_1b_2 sin\alpha)\vert 0_D0_A \rangle+(a_1a_2 sin\alpha + a_1b_2 cos\alpha)\vert0_D1_A \rangle 
\pm &[b_1a_2\vert1_D0_A\rangle\ + b_1b_2\vert1_D1_A\rangle].
\end{eqnarray}
For successful discrimination of data qubit states $\vert \psi_1 \rangle$ and $\vert \psi_2 \rangle$, the condition on the coefficients of Eqn. (11) is,
\begin{eqnarray}
(a_1a_2 cos\alpha-a_1b_2 sin\alpha)=b_1a_2,
\end{eqnarray}
 This yields,
\begin{eqnarray}
 U\vert\psi_{DA}\rangle=[b_1a_2\vert 0_D0_A \rangle &+&(a_1a_2 sin\alpha + a_1b_2 cos\alpha)\vert0_D1_A \rangle] \cr
&\pm &[b_1a_2\vert1_D0_A\rangle\ + b_1b_2\vert1_D1_A\rangle].
\end{eqnarray}
Equation (13) can be re-written as,
\begin{eqnarray}
U\vert\psi_{DA}\rangle=\sqrt2 b_1a_2\vert \pm \rangle+(a_1a_2 sin\alpha+a_1b_2 cos\alpha) \vert0_D1_A \rangle \pm b_1b_2\vert1_D1_A\rangle,
\end{eqnarray}
where $\vert\pm\rangle=1/\sqrt2(\vert0_D0_A\rangle\pm\vert1_D0_A\rangle)$.

 If the final 
state (eqn. 14) contains state $\vert+\rangle$ then the initial state of the data qubit is $\vert \psi_1 \rangle$ if the final state contains state 
$\vert-\rangle$ then 
the initial state of the data qubit is $\vert \psi_2 \rangle$. Square of the coefficient ($\sqrt2 b_1 a_2$) of the state $\vert\pm\rangle$ is called the probability (P) 
of discrimination \cite{dus}.
   
   Equation (12) gives the condition on the ancilla qubit state. For example, when  $\alpha=90^o$, $(b_2/a_2)=-(b_1/a_1)$ 
, and from Eqn.s 2 and 6 it is seen that $\vert\psi_A\rangle$=$\vert\psi_2\rangle$. However for other values of $\alpha$, $\vert\psi_A\rangle$ differs from $\vert\psi_1\rangle$ 
and $\vert\psi_2\rangle$. Hence $\alpha=90^o$ is a special case of Eqn. (12).

   In NMR, the measurement is performed on an ensemble and the results are contained in the expectation values  ($\langle \sigma_x \rangle$
and $\langle \sigma_y \rangle$) of Pauli spin matrices, which in the frequency space yield intensities of various transitions. From Eqn. (14) it is noted that the two transitions of the data qubit 
have different intensities. The $\vert 0_D 0_A \rangle \leftrightarrow \vert 1_D 0_A \rangle$ transition has the  intensity $\pm b_1^2 a_2^2$,  and 
the $\vert 0_D 1_A \rangle \leftrightarrow \vert 1_D 1_A \rangle$ transition has the intensity $\pm (a_1a_2 sin(\alpha)+a_1b_2 cos(\alpha))b_1b_2$. 
To find whether the final state (eqn. 14) contains $\vert+\rangle$ or $\vert-\rangle$, one has to do the projective measurement on the state 
$\vert\pm\rangle$. To simulate projective measurement, we use a method given by Collins for an expectation value quantum search
\cite{col,kim}.
In our experiments, the goal of the projective measurement is to collapse the state of the ancilla qubit (given by equation 14) to
$\vert0_A\rangle$ so that the data qubit gives only one peak which is the coherence of the superposition state $\vert\pm\rangle$ (or 
$\vert 0_D 0_A \rangle \leftrightarrow \vert 1_D 0_A \rangle$ transition).
One can collapse the state of the ancilla qubit to $\vert0_A\rangle$
 by adding two experiments (detection on the data qubit), one without controlled-$\sigma_z$ gate and one with
controlled-$\sigma_z$ gate (fig. 2). Here the controlled-$\sigma_z$ gate is given by the 
unitary transformation,

\begin{eqnarray}
\sigma_z^c=\pmatrix{ 1&0&0&0 \cr 0&1&0&0 \cr 0&0&1&0 \cr 0&0&0&-1 }.
\end{eqnarray}

Controlled-$\sigma_z$ gate inverts the sign of the state$\vert1_D1_A\rangle$.
When $\sigma_z^c$ is applied after the unitary transformation U (eqn. 8), the final state (eqn. 14) becomes,
\begin{eqnarray}
\sigma_z^c U\vert\psi_{DA}\rangle=\sqrt2 b_1a_2\vert \pm \rangle+(a_1a_2 sin\alpha+a_1b_2 cos\alpha) \vert0_D1_A \rangle \mp b_1b_2\vert1_D1_A\rangle.
\end{eqnarray}
 In the first experiment (eqn. 14), intensities of the data qubit transitions corresponding to $\vert 0_D 0_A \rangle \leftrightarrow \vert 1_D 0_A \rangle$
 and $\vert 0_D 1_A \rangle \leftrightarrow \vert 1_D 1_A \rangle$ are, $\pm b_1^2 a_2^2$ and $\pm (a_1a_2 sin(\alpha)+a_1b_2 cos(\alpha))b_1b_2$ 
respectively, whereas in the second experiment (eqn. 16) intensities are  $\pm b_1^2 a_2^2$ and $\mp (a_1a_2 sin(\alpha)+a_1b_2 cos(\alpha))b_1b_2$ respectively.
Thus when the two experiments are added, intensity of $\vert 0_D 1_A \rangle \leftrightarrow \vert 1_D 1_A \rangle$ transition goes to zero and that of 
$\vert 0_D 0_A \rangle \leftrightarrow \vert 1_D 0_A \rangle$ to $\pm 2b_1^2 a_2^2$.  Hence by the above procedure the 
ancilla qubit state is collapsed to $\vert 0_A \rangle$, and the phase 
of the observed transition yields the result of the measurement. 
If the phase is positive then the data qubit is $\vert \psi_1 \rangle$, and 
if it is negative then the data qubit is $\vert \psi_2 \rangle$. The resultant intensity $(2b_1^2a_2^2)$ gives the probability of successful discrimination.

   For the case of linearly polarized states (eqn. 2, y=0; $a_1 = cos\theta_1$, $b_1 = sin\theta_1$), the data qubit states  $\vert \psi_1 \rangle$ and $\vert \psi_2 \rangle$ are schematically shown in Fig. (1b), and $\eta$ and $\phi$ of Eqn. (5) are respectively given by 
$\theta_1$ and zero. In this case ancilla qubit state
(fig. 1b)  
is given by Eqn. (6),  with $a_2=cos\theta_2$ and $b_2=sin\theta_2$. Rest of the procedure to discriminate $\vert \psi_1 \rangle$ and 
$\vert \psi_2 \rangle$ 
remains the same and the probability of discrimination is given by $2b_1^2a_2^2$.

\section{III. Experiment}
In  NMR  spin-1/2 nuclei having sufficiently different Larmor frequencies and weakly coupled to each other by indirect exchange
(J) couplings are used as qubits. 
 The Hamiltonian of the two weakly coupled spin-1/2 nuclei is of the form,
\begin{eqnarray}
H=\omega_1 I_{z1}+\omega_2 I_{z2}+2\pi J_{12} I_{z1}I_{z2}.
\end{eqnarray} 
We have used a Carbon-13 labeled $^{13}CH Cl_3$ as a two qubit system, where the proton ($^{1}H$) and the labeled carbon($^{13}C$) act as two individual 
qubits. J-coupling between $^{13}C$ and $^{1}H$ is 209 Hz. The measured longitudinal ($T_1$) and transverse ($T_2$) relaxation times of $^{1}H$ and 
$^{13}C$ are: $^{1}H$
($T_1$=4.8s and $T_2$=3.3s), and $^{13}C$ ($T_1$=17.2s and $T_2$=0.35s).
 To implement the circuit of Fig. (2), the data ($^{1}H$) and ancilla ($^{13}C$) qubits
have to be first prepared in a pure state. However in NMR pure states are difficult to prepare, instead we prepare pseudopure states which mimics the 
pure states. Several methods are known for the preparation of pseudopure states \cite{cor,pps1,pps2,pps3,pps4,kd,ts}. Here we use spatial averaging 
method \cite{jdu} to prepare pseudopure state using the pulse sequence given in Fig. (3). This pulse sequence \cite{jdu} is specific to labeled 
$^{13}C$-$^{1}H$ 
system and different from homo nuclear case. The details of the preparation of pseudopure state are given in figure captions. Spectra of equilibrium 
state and pseudopure state are shown in Fig. (4). After preparation of pseudopure state, the quantum circuit of Fig. (2)  is implemented by the pulse 
sequence given in Fig. (5).
The pulse sequence in Fig. (5) consists of three  parts,

  (i) Preparation of initial state  ($\vert \psi_{DA}\rangle$): After preparation of pseudopure state($\vert 0_D0_A \rangle$), the data qubit ($^{1}H$) 
is prepared in elliptically polarized state $\vert \psi_D\rangle$ (eqn. 5) by applying a $2\eta$ pulse of appropriate phase on the data 
qubit state $\vert 0_D \rangle$. To prepare the data qubit in state $\vert \psi_1 \rangle$ or $\vert \psi_2 \rangle$ , the phase of the $2\eta$ pulse 
is ($\pi/2+\phi$) or 
$-(\pi/2+\phi)$ respectively (Appendix). 
The ancilla qubit 
($^{13}C$) is prepared in state $\vert \psi_A \rangle$ (eqn. 6) by using Eqn. (12). For example for $\alpha=90^o$, since 
$\vert \psi_A\rangle$=$\vert \psi_2\rangle$, the ancilla qubit  
$\vert \psi_A \rangle$ is prepared by another $(2\eta)_{-(\pi/2+\phi)}$ pulse. For arbitrary $\alpha$, the ancilla qubit is prepared using 
Eqn. (12) with appropriate pulse angle 
and phase. In case of linearly polarized state (eqn. 2, y = 0), $\eta$=$\theta_1$ and $\phi=0^o$, the data qubit can be prepared in states 
$\vert \psi_1 \rangle$ and 
$\vert \psi_2\rangle $ respectively by applying $(2\theta_1)_y$ and $(2\theta_1)_{-y}$ pulses (Appendix) on $\vert 0_D \rangle$. The 
ancilla qubit 
(eqn. 6) is then prepared by applying $(2\theta_2)_y$ pulse on $\vert 0_A \rangle$, where $2\theta_2$ 
is calculated according to Eqn. (12). From Eqn. (12), $2\theta_2$ can take positive as well as negative values. For example When 
$\alpha$= $60^o$,
and $2\theta_1$= $20^o$, $40^o$, $60^o$, $80^o$, $90^o$, $100^o$, $120^o$, $140^o$, $160^o$, 
$2\theta_2$ takes the values, $41^o$, $17.8^o$, $-10.2^o$, $-42.8^o$, $-60^o$, $-77.2^o$, $-109.8^o$,$-137.8^o$, $-161^o$ 
respectively. Here one should note that $(-2\theta_2)_y$ pulse is identical 
to $(2\theta_2)_{-y}$ pulse.

  (ii) Applying unitary operator U : The unitary operator U (fig. 2) is prepared by using two CNOT gates, two NOT gates and two other single qubit gates 
$u_1$ and $u_2$ (fig. 5). The NOT gates on the data qubit ($^{1}H$) are implemented by $(\pi)_x$ pulse and $u_1$, $u_2$ on ancilla qubit by $(\alpha)_{-y}$ and 
$(\alpha)_y$ pulses respectively on $^{13}C$. The CNOT gate is implemented by using the pulse sequence 
$(\pi/2)_z^1$-$(\pi/2)_y^2$-$(1/2J)$-$(\pi/2)_x^2$-$(\pi/2)_{-z}^2$ \cite{cor}, where the superscript 1 stands for proton and 2 stands for carbon. The $(\pi/2)_z^1$ 
is obtained by the composite pulse  $(\pi/2)_y^1$-$(\pi/2)_x^1$-$(\pi/2)_{-y}^1$ as shown in Fig. 5. The $(\pi/2)_{-z}^2$ pulse can be obtained by a 
another composite pulse  
$(\pi/2)_{-x}^2$-$(\pi/2)_y^2$-$(\pi/2)_y^2$, so that the first $(\pi/2)_{-x}^2$ pulse of the composite pulse cancels the last $(\pi/2)_x^2$ pulse of 
the CNOT gate yielding the last two pulses in the CNOT sequence as $(\pi/2)_{-y}^2$-$(\pi/2)_x^2$. All the pulses in the pulse sequence are applied at resonance, so the chemical shifts are 
refocused throughout the pulse sequence. Hence during the time period (1/2J), system evolves only under the J-coupling Hamiltonian 
$H_J= 2\pi J I_{z1}I_{z2} $ yielding the unitary operator, $e^{-i\pi I_{z1}I_{z2}}$.

  (iii) Controlled-$\sigma_z$ gate ($\sigma_z^c$) is implemented by $(\pi/2)_z^{1,2}$ pulse followed by an evolution for the time 1/2J
\cite{nmr7}. 
$(\pi/2)_z^{1,2}$ 
pulses are realized by composite rotation on both qubits as shown in Fig. 5.

\begin{center}
(a) Linearly polarized states:
\end{center}

 We have studied the linearly polarized case by varying both the parameters $\alpha$ (rotation angle of U, eqn. 8) and    
$2\theta_1$ (angle between $\vert \psi_1 \rangle$ and $\vert \psi_2 \rangle$, fig. 1b). The pulse sequence given in Fig. (5) is implemented, with the 
initial state  prepared 
as described above (in section III(i)). Experiment is performed to discriminate several 
pairs of linearly polarized states for $\alpha$=$30^o$, $45^o$, $60^o$, and $90^o$.
For each value of $\alpha$, the experiment is carried out for  $2\theta_1$=$20^o$, $40^o$, $60^o$, $80^o$, $90^o$, $100^o$, $120^o$, $140^o$, $160^o$. As 
mentioned in 
theory section (II), the experiment is performed twice, one with $\sigma_z^c$ and other without $\sigma_z^c$, and the results are added so that the resultant 
intensity of the data qubit transition gives the probability of discrimination (P=$2b_1^2a_2^2$). Figure (6) contains typical spectra for 
$2\theta_1$= $90^o$ and 
$\alpha$ = $90^o$, $60^o$, $45^o$, $30^o$, where the data qubit is prepared respectively in states $\vert \psi_1 \rangle$ (fig. 6a-d) and 
$\vert \psi_2 \rangle$ (fig. 6e-h). As 
shown in Fig. (6), the positive intensities of the resultant peaks indicate that the initial state of data qubit is $\vert \psi_1\rangle$ and the negative 
intensities of the resultant peaks indicate that the initial state of data qubit is $\vert \psi_2 \rangle$. The intensity of the peak 
yields the probability (P=$2b_1^2a_2^2$). In Fig. (6) one can observe that the intensity of the resultant peak (probability of 
 discrimination) changes with $\alpha$.
For different values of $\alpha$, the probability of discrimination P (experimental and simulation results) as a function of $2\theta_1$ 
is given 
in Fig. (7). From Fig. 7 one can find the optimum angle $2\theta_1$ for maximum probability of discrimination for a given value of $\alpha$. Figure (8), 
on the other hand, shows the variation of Probability of discrimination(P) as a function of $\alpha$, for different 
$2\theta_1$.
From Fig. (8), one can find the value of $\alpha$ to get the maximum probability of discrimination for a given angle ($2\theta_1$) between 
$\vert \psi_1 \rangle$ and $\vert \psi_2 \rangle$. In both figures 7 and 8, the experimental points agree well with the simulations, confirming 
successful discrimination of linearly 
polarized states $\vert \psi_1 \rangle$ and $\vert \psi_2 \rangle$ of the data qubit. 

\begin{center}
(b) Elliptically polarized states:
\end{center}
  We also discriminate several pairs of elliptically polarized states. Experiments have been performed, using the 
pulse sequence given in Fig. (5), for $\alpha$=$90^o$ and ellipticities $\epsilon$= $0^o$,$15^o$, and $30^o$. For each value of $\epsilon$, 
we perform the 
experiment for $2\theta_1$=$20^o$, $40^o$, $60^o$, $90^o$, $120^o$, $140^o$, $160^o$. As described above (in section III(i)), the data qubit states 
$\vert \psi_1 \rangle$ or $\vert \psi_2 \rangle$ of Eqn. (5) are prepared respectively by applying a $(2\eta)_{(\pi/2 + \phi)}$ or a  
$(2\eta)_{-(\pi/2 +\phi)}$ 
pulse, where 
$\eta$ and $\phi$ are calculated from Eqn. (4). Ancilla qubit is prepared by using Eqn.(12). For $\alpha$=$90^o$, since 
$\vert \psi_A \rangle$=$\vert \psi_2 \rangle$, ancilla qubit is prepared by $(2\eta)_{-(\pi/2 + \phi)}$ pulse.
Figure (9) shows both experimental and  simulated results of the probability of  successful discrimination
of pair of elliptically polarized states $\vert \psi_1 \rangle$ and $\vert \psi_2\rangle$ as a function of $2\theta_1$ 
(the angle between $\vert \psi_1 \rangle$ and 
$\vert \psi_2 \rangle$, as shown in fig. 1a), for different ellipticities for a fixed value of $\alpha$=$90^o$. From Fig. (9) one can obtain the probability of 
discrimination (P) of pair of elliptically polarized states, as a function of ellipticity. However for $\alpha$=$90^o$ the maximum probability of
 discrimination for any ellipticity is always obtained for $2\theta_1$= $90^o$. The experimental results for low ellipticities (fig. 9) 
match  
well with the theoretical results, but deviates for higher ellipticities. Similar results have been obtained in 
optics, where partial measurement in the Bell basis has been done for the discrimination of elliptically polarized states. \cite{dus1}. 

 \section{IV. conclusion}
 The implementation of a programmable quantum state discriminator by NMR has been demonstrated. The device  
discriminates pair of data qubit states unambiguously (error free) that are symmetrically located around some fixed state. One can use the 
same device (without changing it's parameters) to discriminate any pair of data qubit states, by suitably preparing the ancilla qubit. 
However the probability of discrimination depends on the parameter of the device (angle $\alpha$).   
It may be noted that since NMR is an ensemble measurement, it is inevitable that to do projective measurement one has to prepare the input 
state twice.
The probability of successful
discrimination is obtained as a function of the angle between pair of data qubit states and the rotation angle of the unitary operator of the protocol. 
 The states of the ancilla (programme) qubit that represent different programs can be nonorthogonal, which indicates the 
quantum nature of the programming. It is further shown that if the pair of data qubits are in elliptically polarized states then the probability of successful 
discrimination is also a function of ellipticity. 
    
\section{acknowledgment}
Useful discussions with Arindam Ghosh and Karthick Kumar are gratefully acknowledged.
The use of DRX-500 NMR spectrometer funded by the Department of
Science and Technology (DST), New Delhi, at the Sophisticated Instruments Facility, Indian Institute of Science,
Bangalore, is gratefully acknowledged. AK acknowledges "DAE-BRNS" for the award of "Senior Scientists scheme",
and DST for a research grant on "Quantum Computing using NMR techniques".

\section{appendix}
Unitary operator corresponding to a radio frequency (r.f) pulse of angle $\alpha$ and phase (direction of r.f pulse) $\phi$ is, 
$R_{\phi}(\alpha)$, which is also called as $(\alpha)_\phi$ pulse,

 \hspace{6cm} $R_{\phi}(\alpha)=e^{-i\alpha \hat{n}.I}$,

 where $\hat{n}$ is a unit vector whose direction is along the direction of r.f pulse and $I=I_x\hat{i}+I_y\hat{j}+I_z\hat{k}$, where I is
the angular momentum operator of spin 1/2 nuclei. 

  In spherical polar coordinates $\hat{n}$=$I_xcos(\phi)sin(\theta)\hat{i}+I_ysin(\phi)sin(\theta)\hat{j}+I_zcos(\theta)\hat{k}$, where
$\theta$ is the angle between $\hat{n}$ and z-axis (direction of static magnetic field), and $\phi$ is the angle between $\hat{n}$ and x-axis.
Here $\theta=90^o$, since r.f pulse is applied perpendicular to static magnetic field.
 
  After simplification, unitary operator of $(\alpha)_\phi$ pulse, $R_{\phi}(\alpha)$ can be written as,

     \vspace{0.5cm} $R_{\phi}(\alpha)=\pmatrix{ cos(\alpha/2)&-e^{-i\phi_1}sin(\alpha/2) \cr e^{i\phi_1}sin(\alpha/2)&cos(\alpha/2)}$, 
\vspace{0.5cm}
where $\phi_1=\phi-\pi/2$.

 \vspace{0.5cm} Here $\phi=0^o$ gives $(\alpha)_x$ pulse, and $\phi=180^o$ gives $(\alpha)_{-x}$ pulse. Similarly $\phi=90^o$ and $\phi=270^o$ gives 
$(\alpha)_y$ and $(\alpha)_{-y}$ pulses respectively. From $R_{\phi}(\alpha)$, one can calculate any unitary operator, corresponding to any 
arbitrary angle and phase.
For example the unitary operator corresponding to $(2\eta)_{ (\pi/2+\phi)}$ pulse is,
    
     \vspace{0.5cm}  $(2\eta)_{(\pi/2+\phi)}=\pmatrix{ cos(\eta)&-e^{-i\phi}sin(\eta) \cr e^{i\phi}sin(\eta)&cos(\eta)}$. 
  
   \vspace{0.5cm} The unitary operator of $(2\eta)_{-(\pi/2+\phi)}$ pulse is given by the Hermitian conjugate of the above.

\pagebreak

\pagebreak

\hspace{6cm}\large{FIGURE CAPTIONS} :

 FIG. 1: (a) Pictorial representation of elliptically polarized states $\vert \psi_1 \rangle$ and $\vert \psi_2 \rangle$ of the data qubit. They are 
symmetrically placed with respect to $\vert 0 \rangle$.  When $2\theta_1$=$90^o$ the two states are orthogonal. Ellipticity $\epsilon$ is defined as, 
$\epsilon = y/x$. When 
y=0, $\epsilon=0^o$, $\vert \psi_1 \rangle$ and $\vert \psi_2 \rangle$ are linearly polarized states. 
         
         (b) Pictorial representation of linearly polarized states $\vert \psi_1 \rangle$ and $\vert \psi_2 \rangle$ of the data qubit, 
$\vert \psi_A \rangle$ is the ancilla qubit. When the data 
qubits ($\vert \psi_1 \rangle$ and $\vert \psi_2 \rangle$) are elliptically polarized states as shown in Fig. (1a), then ancilla qubit 
$\vert \psi_A \rangle$ is also elliptically polarized state (not shown in fig. (1a)).
 
 FIG. 2:  Quantum circuit for the discrimination of data qubit state 
$\vert \psi_D \rangle$= $\vert \psi_1 \rangle$ or $\vert \psi_2 \rangle$, using an ancilla qubit, prepared in state $\vert \psi_A \rangle$.
The unitary operator U needed for such a protocol consists of two CNOT gates, two NOT gates (X) and two other single qubit gates $u_1$ and $u_2$. For projective measurement
controlled-$\sigma_z$ gate is needed at the end of the of the protocol, as described in the text.

 FIG. 3: The pulse sequence for creation of pseudopure state from the equilibrium state for a proton - carbon-13 two qubit system, using the method of 
spatial averaging \cite{jdu}. In the product 
operator formalism\cite{ern}, equilibrium magnetization can be represented by $4I_{1z}$+$I_{2z}$, where 1 stands for proton and 2 stands for carbon 
(since $\gamma_{^{1}H} \simeq 4 \gamma_{^{13}C}$). All the pulses 
are applied on $^{1}H$ so there is no change in carbon magnetization. $4I_{1z}$ is converted to 2($I_{1z}$-$\sqrt(3)I_{1y}$) by $(\pi/3)_x$ pulse, 
and gradient pulse kills the transverse magnetization $2\sqrt(3)I_{1y}$. The remaining magnetization of $^{1}H$, $2I_{1z}$ is converted to 
$\sqrt(2)(I_{1z}-I_{1y})$, by 
a $(\pi/4)_x$ pulse. Evolution under J-coupling for time 1/2J (i.e. evolution under the unitary operator $e^{-i\pi I_{1z}I_{2z}}$) converts it to 
$\sqrt(2)(I_{1z}+2I_{1x}I_{2z})$, which is converted to $(I_{1z}-I_{1x})+(2I_{1x}I_{2z}+2I_{1z}I_{2z})$ by a $(\pi/4)_{-y}$ pulse. At the end a gradient pulse is applied 
to kill the transverse magnetization, yielding the magnetization ($I_{1z}$+$I_{2z}$+$2I_{1z}I_{2z}$), which is a $\vert 00 \rangle$ pseudopure state. All the pulses are applied at resonance so that chemical shifts are refocused through out the pulse sequence.

FIG. 4: (a) Equilibrium $^{1}H$ and $^{13}C$ spectra of $^{13}CHCl_3$ dissolved in $CDCl_3$.

         (b) spectra obtained after the preparation of pseudopure state by using the method of spatial averaging using the pulse 
          sequence given in Fig. (3). To obtain these spectra, $\pi/2$ read pulses are used on each spin. The appearance of a single resonance
line with positive intensity for each spin (double the intensity of carbon and half that of proton compared to respective equilibrium 
spectra (fig. 4a)),
is a confirmation of the $\vert 00 \rangle$ pseudo pure state ($I_{1z}$+$I_{2z}$+$2I_{1z}I_{2z}$). 

 FIG. 5:  The pulse sequence for implementation of the quantum circuit of Fig. 2. The data qubit ($^{1}H$) is prepared in elliptically 
polarized 
states $\vert \psi_1 \rangle$ and $\vert \psi_2 \rangle$ (eqn. 5) by $(2\eta)_{(\pi/2+\phi)}$ and $(2\eta)_{-(\pi/2+\phi)}$ pulses respectively 
and ancilla 
qubit ($^{13}C$) is prepared in state $\vert \psi_A \rangle$, by $(2\eta)_{-(\pi/2+\phi)}$ pulse for $\alpha=90^o$. In case of linearly polarized 
states (eqn. 2, y=0),
data qubit is prepared in states $\vert \psi_1 \rangle$ and $\vert \psi_2 \rangle$ by $(2\theta_1)_{y}$ and $(2\theta_1)_{-y}$ pulses respectively 
and ancilla 
qubit is prepared in state $\vert \psi_A \rangle$  by $(2\theta_2)_{y}$ pulse, 
where $2\theta_2$ is calculated according
to Eqn. (12). Figure (2) contains four single qubit gates and two CNOT gates. NOT 
gate (represented by X in fig. 2, eqn. 9) is implemented by $\pi_x$ pulse. $u_1$ and $u_2$ (eqn. 9) are implemented by 
$(\alpha)_{-y}$ and $(\alpha)_y$ 
pulses respectively. CNOT gate (eqn. 10) is implemented by the pulse sequence 
$(\pi/2)_z^1$-$(\pi/2)_y^2$-$(1/2J)$-$(\pi/2)_x^2$-$(\pi/2)_{-z}^2$, where $(\pi/2)_z^1$ pulse is obtained by the composite 
pulse $(\pi/2)_y^1$-$(\pi/2)_x^1$-$(\pi/2)_{-y}^1$ and $(\pi/2)_{-z}^2$ pulse is obtained by the composite pulse 
$(\pi/2)_{-x}^2$-$(\pi/2)_y^2$-$(\pi/2)_{x}^2$. The first $(\pi/2)_{-x}^2$ pulse of composite (-z) pulse is canceled with the last 
$(\pi/2)_{x}^2$
 pulse of CNOT gate. All the pulses are applied at resonance, such that the chemical shifts are  refocused 
throughout the pulse sequence. 

     FIG. 6:  Proton spectra of $^{13}CHCl_3$ obtained after the implementation of pulse sequence given in Fig. (5), where the initial 
states of data and ancilla qubit are prepared in linearly polarized states (section III(i)), 

 (a) $\vert \psi_D \rangle$=$\vert \psi_1 \rangle$, $2\theta_1$=$90^o$, and $\alpha$=$90^o$

 (b) $\vert \psi_D \rangle$=$\vert \psi_1 \rangle$, $2\theta_1$=$90^o$, and $\alpha$=$60^o$

 (c) $\vert \psi_D \rangle$=$\vert \psi_1 \rangle$, $2\theta_1$=$90^o$, and $\alpha$=$45^o$

 (d) $\vert \psi_D \rangle$=$\vert \psi_1 \rangle$, $2\theta_1$=$90^o$, and $\alpha$=$30^o$

 (e) $\vert \psi_D \rangle$=$\vert \psi_2 \rangle$, $2\theta_1$=$90^o$, and $\alpha$=$90^o$

 (f) $\vert \psi_D \rangle$=$\vert \psi_2 \rangle$, $2\theta_1$=$90^o$, and $\alpha$=$60^o$

 (g) $\vert \psi_D \rangle$=$\vert \psi_2 \rangle$, $2\theta_1$=$90^o$, and $\alpha$=$45^o$

 (h) $\vert \psi_D \rangle$=$\vert \psi_1 \rangle$, $2\theta_1$=$90^o$, and $\alpha$=$30^o$

    In each of the above experiments $\vert \psi_A \rangle$ is initialized by choosing $2\theta_2$ to satisfy Eqn. (12). A complete set of these 
experiments have been carried out for different values of $\alpha$ by varying $2\theta_1$ and $2\theta_2$ (satisfying eqn. 12). The results are plotted 
in Fig.s (7,8).

 FIG. 7: Probability of discrimination P (resultant intensity of the transition of the data qubit, $^{1}H$) from Fig. (6) 
and corresponding experiments 
for various $2\theta_1$ and $\alpha$. The $2\theta_2$ is adjusted to satisfy Eqn. (12) in each case. The expected intensities 
(shown by thick lines) are 
obtained by simulation of the pulse programme of the pulse sequence given in Fig. (5) using MATLAB programme. Since the total pulse 
sequence lasts for about 11.8 ms, which is much less than $T_1$ and $T_2$ of both $^{1}H$ and $^{13}C$, the relaxation effects were not 
included in the simulation. However all the experimental data points are normalized with respect to the experimental spectrum of 
$\alpha = 90^o$ and $2\theta_1=90^o$ for which the intensity is taken as 0.5, which is the theoretical expected intensity. 
The maximum probability of discrimination ($P_{max}$) is obtained for $2\theta_1$=$90^o$ for all values of 
$\alpha$. However, the value of $P_{max}$ depends on the value of $\alpha$.

 FIG. 8: The results shown in Fig. (7), are replotted as a function of $\alpha$ for various $2\theta_1$. The continuous curves are simulated
results and the experimental data points are shown by crosses. From these curves one can find the optimum value of $\alpha$ for a given
$2\theta_1$.

 FIG. 9: Experimental and simulated results of probability of discrimination (P) of pair of elliptically polarized states  $\vert \psi_1 \rangle$ and 
$\vert \psi_2 \rangle$ are shown , as a function of ellipticity ($\epsilon$) and $2\theta_1$ (angle between $\vert \psi_1 \rangle$ and 
$\vert \psi_2 \rangle$) for 
$\alpha$=$90^o$. Simulated results sans relaxation are shown by 
thick lines. However all the experimental spectra are normalized to  $2\theta_1 = 90^o$ for $\epsilon = 0^o$ to the expected value of 0.5.\\

\newpage

\begin{center}
\begin{figure}
\epsfig{file=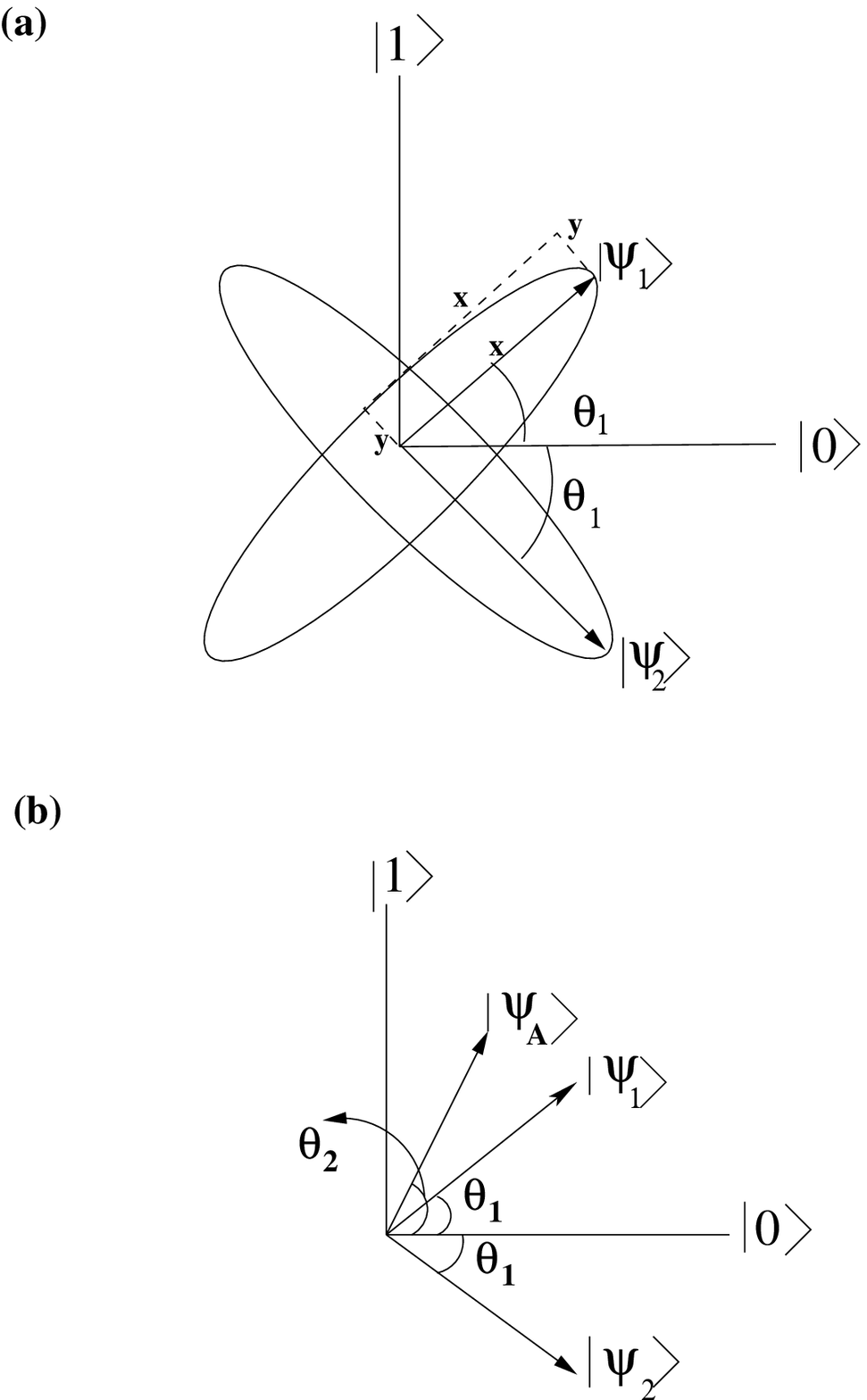,width=10cm}
\caption{}
\end{figure}
\end{center}

\pagebreak
\begin{figure}
\begin{center}
\epsfig{file=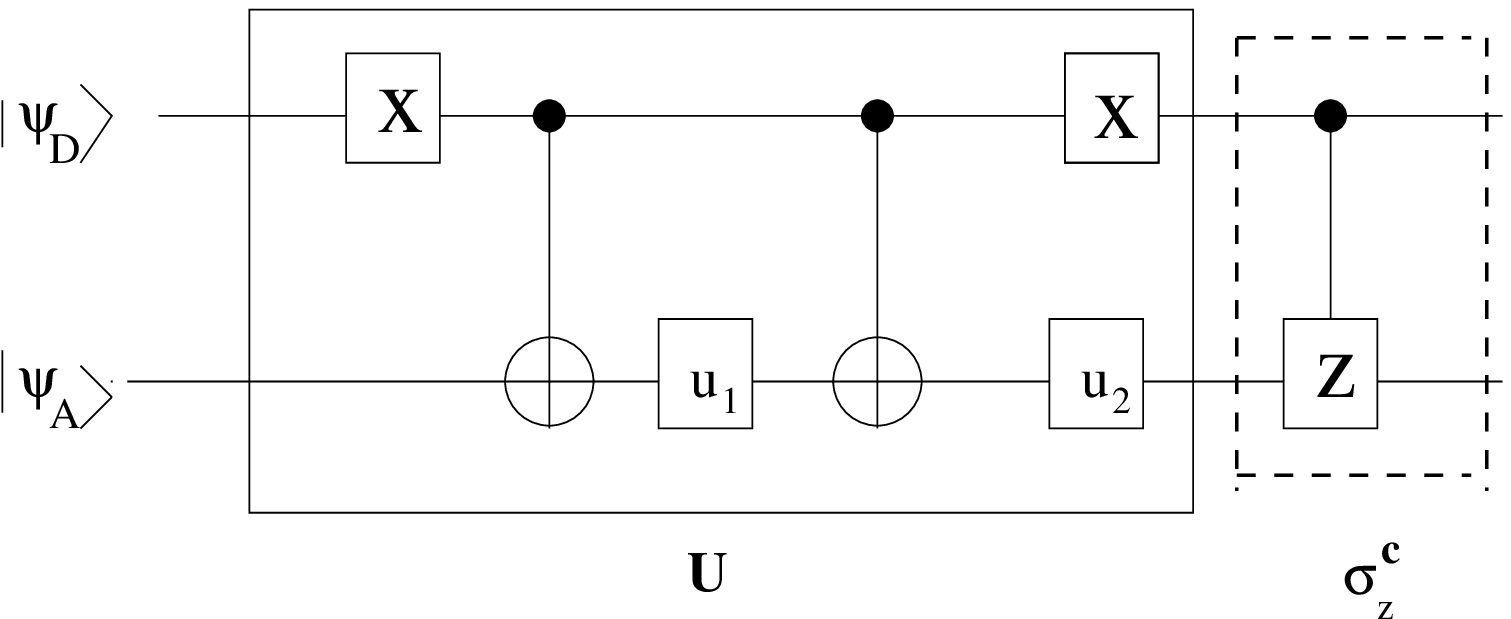,width=13cm}
\caption{}
\end{center}
\end{figure}

\pagebreak
\begin{figure}
\begin{center}
\epsfig{file=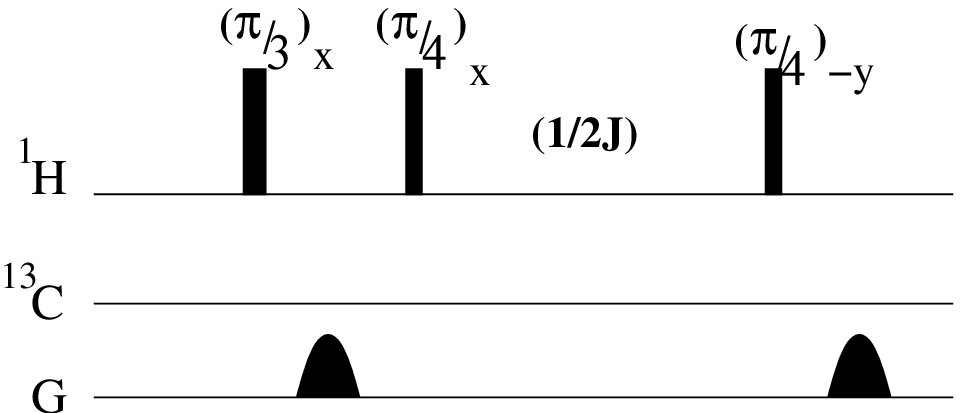,width=10cm}
\caption{}
\end{center}
\end{figure}

\pagebreak
\begin{figure}
\begin{center}
\epsfig{file=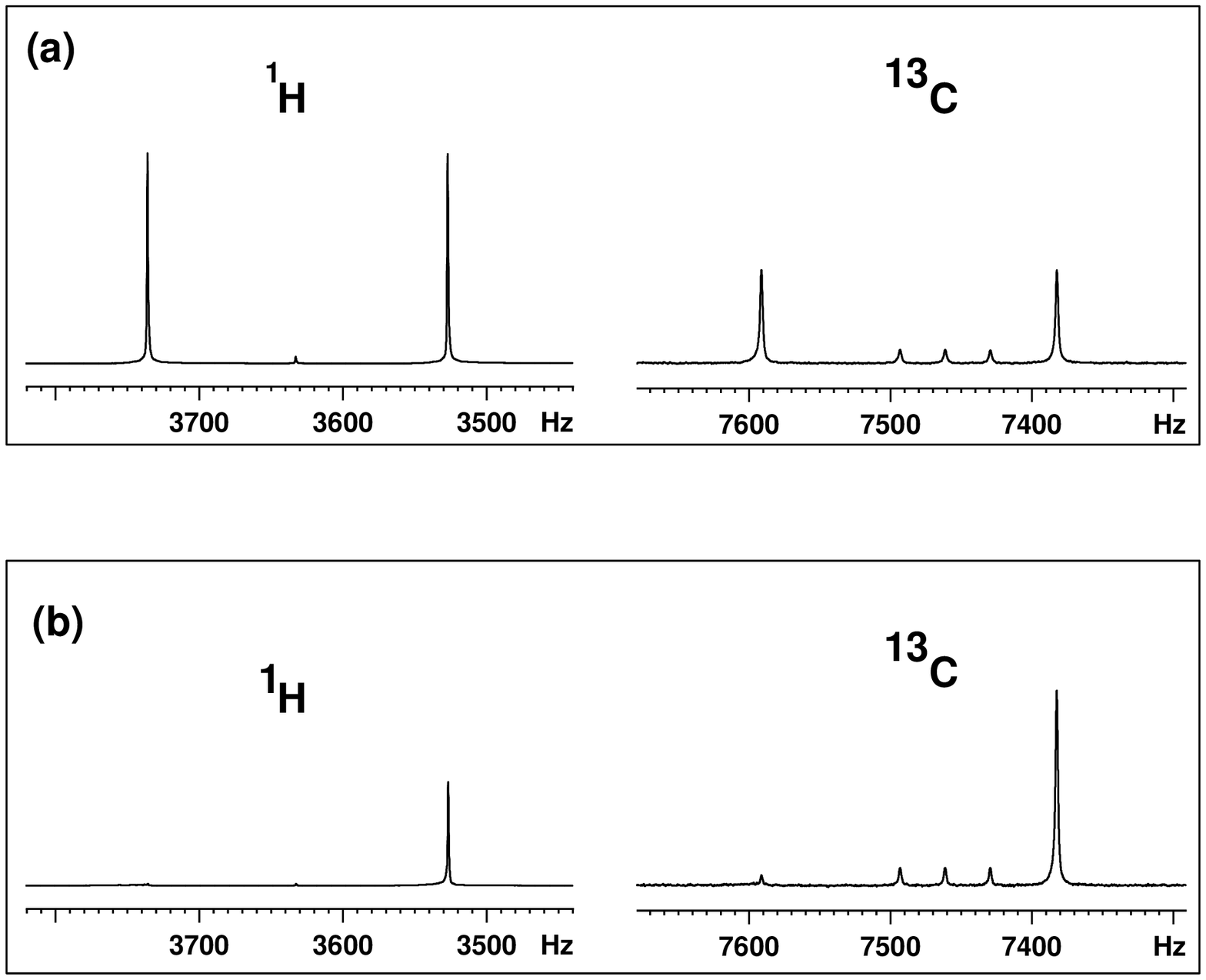,width=10cm}
\caption{}
\end{center}
\end{figure}

\pagebreak
\begin{figure}
\begin{center}
\epsfig{file=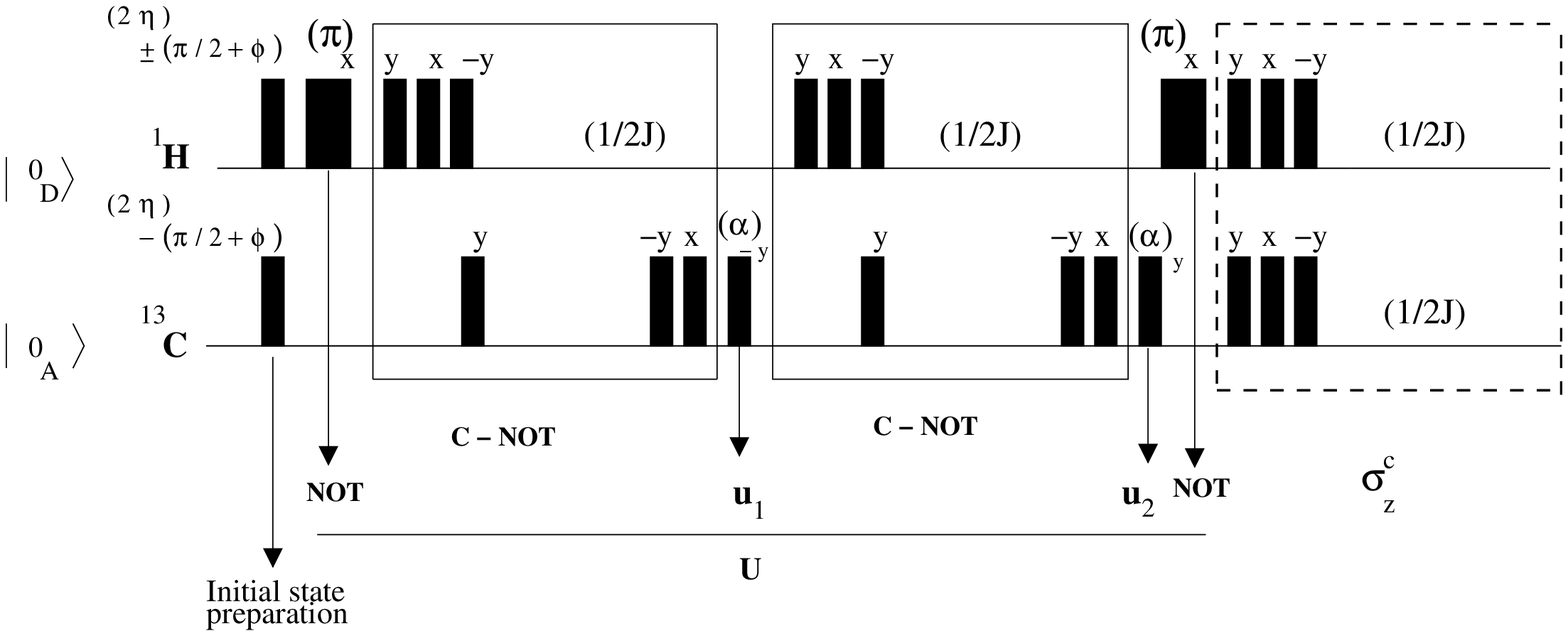,width=15cm}
\caption{}
\end{center}
\end{figure}

\pagebreak
\begin{figure}
\begin{center}
\epsfig{file=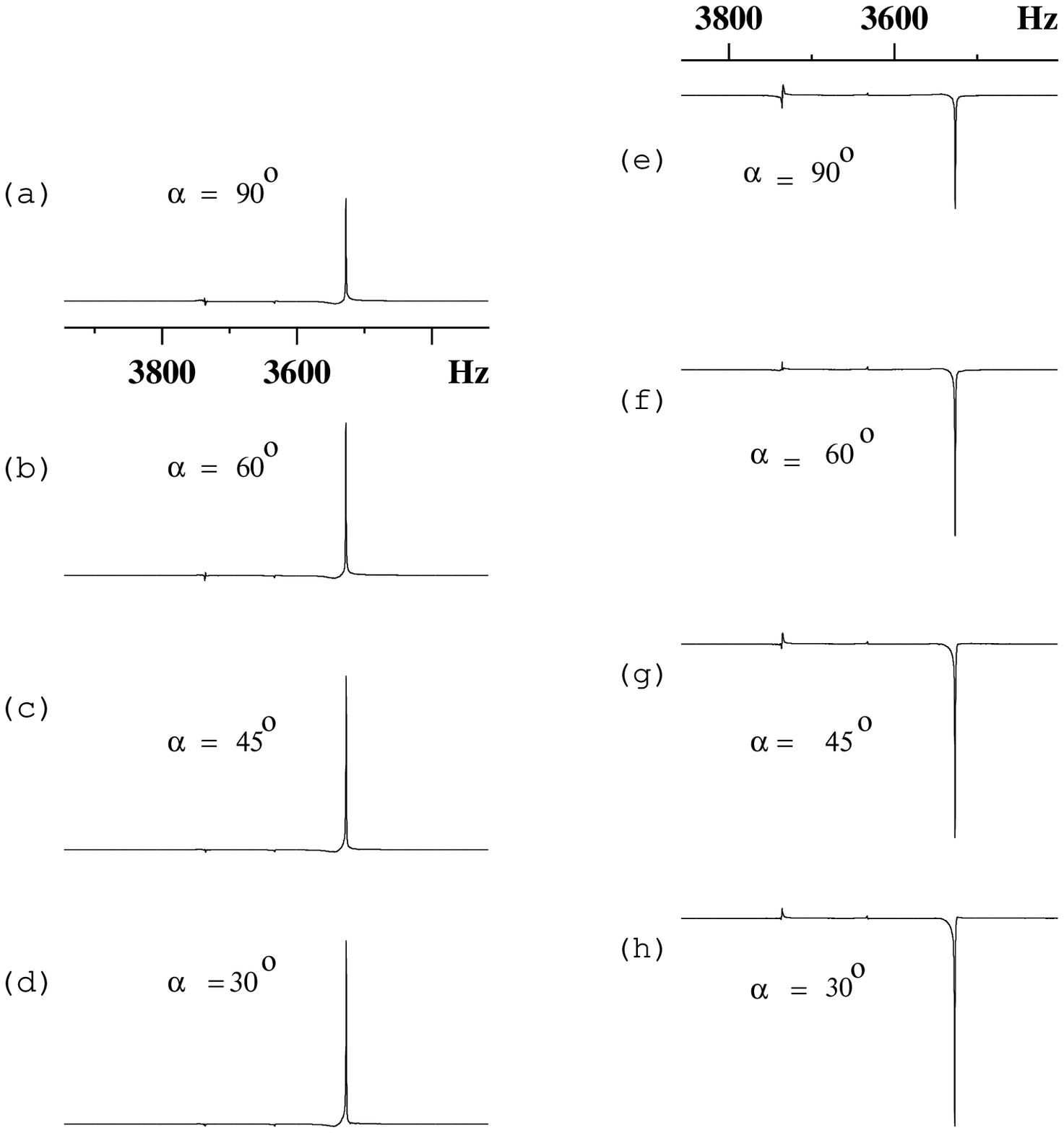,width=10cm,angle=270}
\caption{}
\end{center}
\end{figure}

\pagebreak
\begin{figure}
\begin{center}
\epsfig{file=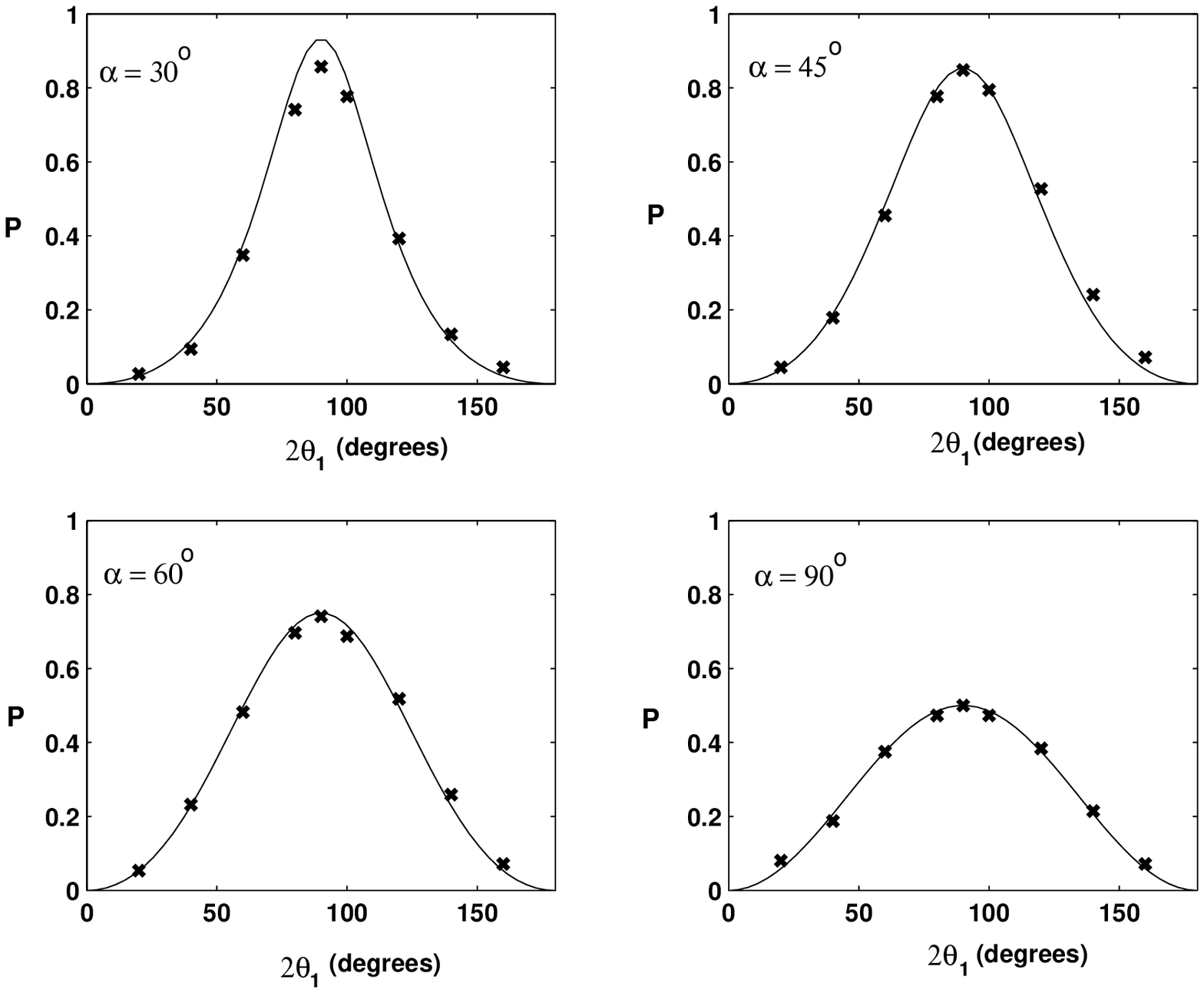,width=15cm}
\caption{}
\end{center}
\end{figure}

\pagebreak
\begin{figure}
\begin{center}
\epsfig{file=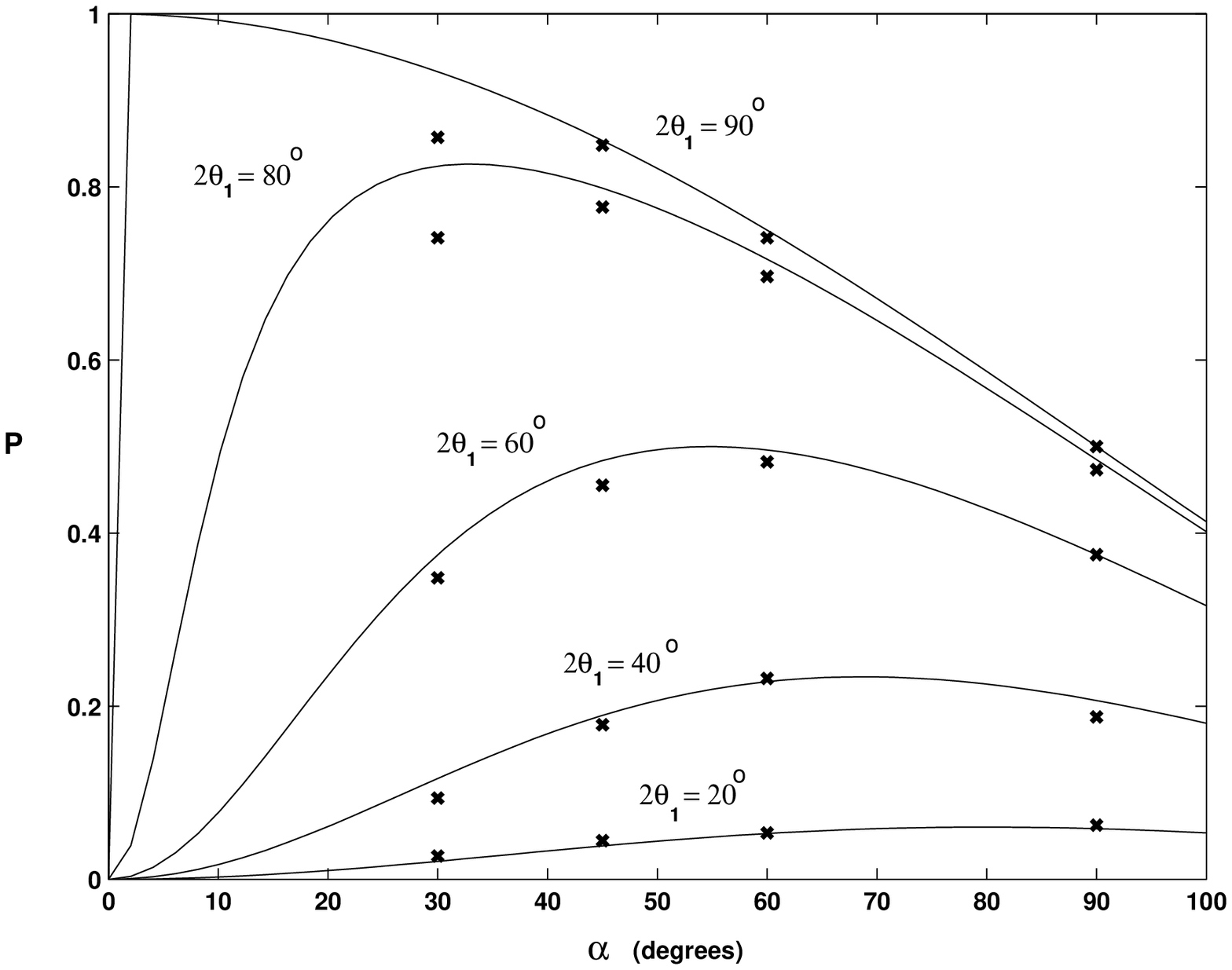,width=15cm}
\caption{}
\end{center}
\end{figure}

\pagebreak
\begin{figure}
\begin{center}
\epsfig{file=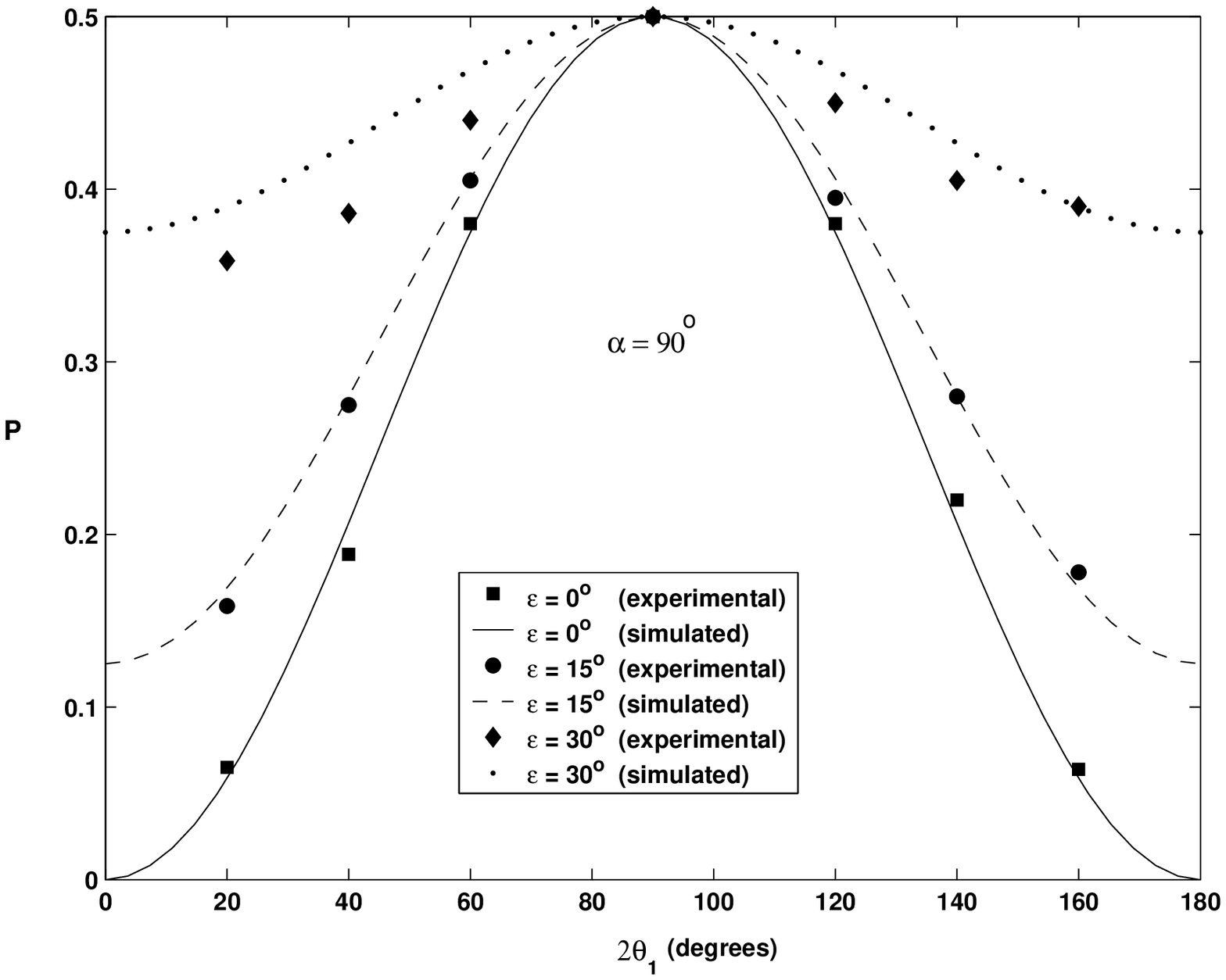,width=15cm}
\caption{}
\end{center}
\end{figure}


\begin{thebibliography}{99}
\bibitem{fey} R. Feynman, {\it Int. j. Theor. phys}. {\bf 21}, 467 (1982).
\bibitem{ben} C.H. Bennett, Int. J. Theor. Phys. 21 905 (1982).
\bibitem{deu} D. Deutsch, Proc. R. Soc. London, Ser. A 400, 97 (1985)
\bibitem{loy} S. Lloyd, Science 261, 1569 (1993).
\bibitem{pw} P. W. Shor, {\it SIAM Rev}. {\bf 41}, 303-332 (1999).
\bibitem{deujoz} D. Deutsch and R. Jozsa, {\it Proc. R. Soc. Lond. A} {\bf 493}, 553 (1992).
\bibitem{gr} L.K. Grover, {\it Phys. Rev. Lett.} {\bf 79}, 325 (1997).
\bibitem{gru} J. Gruska "Quantum Computing", Mcgraw-Hill Limited, UK, 1999.
\bibitem{db} D. Bouwnmeester, A. Ekert, A. Zeilinger(Eds.), "The Physics of
Quantum Information", Springer, Berlin, 2000.
\bibitem{ic} M.A. Nielsen , I.L. Chuang, "Quantum Computation and Quantum
Information". Cambridge University Press, Cambridge, U.K. 2000.
\bibitem{cov} T. M . Cover and J. A. Thomas. Elements of Information Theory. John Wiely and Sons, Newyork, 1991.
\bibitem{nmr1} I. L. chuang, L. M. K. Vanderspyen, X. Zhou, D. W. Leung, and S. Llyod, {\it Nature (london)}, {\bf 393}, 1443 (1998).
\bibitem{nmr2} J.A. Jones and M. Mosca, {\it J. Chem. Phys.} {\bf 109}, 1648 (1998).
\bibitem{nmr3} I.L. Chuang, N. Gershenfeld, M. Kubinec, Phys. Rev. Lett. {\bf 80}, 3408 (1998).
\bibitem{nmr4} J.A. Jones, M. Mosca, and R. H. Hansen, {\it Nature (London)} {\bf 393}, 344 (1998).
\bibitem{nmr5} T. S. Mahesh, Kavita Dorai, Arvind, Anil Kumar, {\it J. Mag. Res.} {\bf 148}, 95 (2001).
\bibitem{nmr6} Neeraj Sinha, T. S. Mahesh, K.V. Ramanathan, and Anil Kumar,
              J. Chem. Phys. 114, (2001) 4415.
\bibitem{nmr7} Ranabir Das, T.S. Mahesh, and Anil Kumar, {\it J. Magn. Reson.} {\bf 159} 46 (2002).
\bibitem{nmr8} Ranabir Das and Anil Kumar, {\it Phys. Rev. A} {\bf 68}, 032304 (2003).
\bibitem{phil} L.S.Philips, S. M. Barnett and D. T. Pegg, {\it Phys. Rev. A}. {\bf 58}, 3259.
\bibitem{zh} Zhang Shengyn,Feng Yuan, Sun Xiaoming, et al. {\it Phys. Rev. A }. {\bf 64}, 062193
\bibitem{barr} S. M. Barnett, {\it Phys. Rev. A}. {\bf 64}, 030303.
\bibitem{che} A.Chefles, {\it Phys. Rev. A}. {\bf 64}, 062305.
\bibitem{che1} A.Chefles, {\it Contemp. Phys}. {\bf 41}, 401 (2001).
\bibitem{hel} C. W. Helsrom, {\it Quantum Detection and Estimation Theory}. (Acadamic Press, New York, 1976).
\bibitem{iva} I. D. Ivanovic, {\it Phys. Lett. A}. {\bf 123}, 257 (1987).
\bibitem{peres} A. Peres, {\it Phys. Lett. A}. {\bf 128}, 19 (1988).
\bibitem{che3} A.Chefles and S. M. Barnett, {\it Phys. Lett. A}. {\bf 250}, 223 (1998).
\bibitem{hut} B. Huttner, A. Muller, J. D. Gautier, H.Zbinden, and N.Gisin {\it Phys. Rev. A}. {\bf 250}, 223 (1998).


\bibitem{dus} Miloslav Dusek and Vladimir Buzek, {\it Phys. Rev. A}. {\bf 66}, 022112 (2002).
\bibitem{fil}  J. Fiurasek, M. Dusek, and R. Filip, {\it Phys. Rev. Lett}. {\bf 89}, 190401 (2002).
\bibitem{fiu}  J. Fiurasek and M. Dusek, {\it Phys. Rev. A}. {\bf 69}, 032302 (2004).
\bibitem{paz}  J. P. Paz and A. Roncaglia, {\it Phys. Rev. A}. {\bf 68}, 052316 (2003).
\bibitem{ari}  G. M. D'Ariano, P. Perinotti, M. F. Sacchi, {\it Europhys. Lett}. {\bf 65}, 165 (2004).


\bibitem{niel} M. A . Nielsen, I.L. Chuang, {\it Phys. Rev. Lett}.{\bf 79}, 321 (1997).
\bibitem{vid}  G. Vidal, L. Masanes, and J. I. Cirac, {\it Phys. Rev. Lett}. {\bf 88}, 047905 (2002).
\bibitem{hil}  M. Hillery, V. Buzek, and M. Ziman, {\it Phys. Rev. A}. {\bf 65}, 022301 (2002).



\bibitem{dus1} Jan Soubusta, Antonin Cernoch, Jaromir Fiurasek, and Miloslav Dusek, {\it Phys. Rev. A}. {\bf 69}, 052321 (2004).
\bibitem{dus2} M. Dusek, M. Jahma, and N. Lutkenhaus, {\it Phys. Rev. A}. {\bf 62}, 022306 (2000).
\bibitem{ham} S.Hamieh, {\it J. Phys. A: Math. Gen.}. {\bf 37}, L 59-L 61 (2004).
\bibitem{yon} Yong Wook Cheong, Hyunjae Kim, and Hai-Woong Lee , {\it Phys. Rev. A}. {\bf 70}, 032327 (2004).
\bibitem{zhu} Zhu - Liang Cao, Wei Song, quant-ph/0401054.
\bibitem{col} D. Collins, {\it Phys. Rev. A} {\bf 65}, 052321 (2002).
\bibitem{kim} Jaehyun Kim, Jae-Seung Lee, Taesoon Hwang, and Soonchil Lee , {\it J. Mag. Res.}, {\bf 166}, 35-38 (2004).
\bibitem{pw1}  P.W. Shor, {\it Phys. Rev. A} {\bf 52}, 2493 (1995).
\bibitem{cor} D.G. Cory et al., Physica D 120, 82 (1998); J. Du etal., phys. rev. lett. 91, 100403 (2003).
\bibitem{pps1} Cory, D. G., Fahmy, A. F. and Havel, T. F., Ensemble quantum computing by NMR spectroscopy. {\it Proc. Natl. Acad. Sci. USA},
              {\bf 94}, 1634 (1997).
\bibitem{pps2} Gershenfeld, N. and Chuang, I. L., Bulk spin-resonance quantum computation. {\it Science}, {\bf 275}, 350 (1997).
\bibitem{pps3} Knill, E., Chuang, I. L. and Laflammem R., Effective pure states for bulk quantum computation, {\it Phys. Rev. A}. {\bf 57}, 3348 (1998). 
\bibitem{pps4} Chuang, I. L., Gershenfeld, N, Kubines. M. G. and Leung, D. W., Bulk quantum computation with nuclear magnetic resonance, 
{\it Proc. R. Soc. London, Ser. A}, {\bf 454}, 447-467 (1998).
\bibitem{kd} Kavita Dorai, Arvind, Anil Kumar, {\it Phys Rev A.} {\bf 61}, (2000) 042306.
\bibitem{ts} T. S. Mahesh, Anil Kumar, Phys. Rev. A {\bf 64}, 012307 (2001).
\bibitem{jdu} J. Du, H. Li, X. Xu, M. Shi, J. Wu, X. Zhou, and R. han, Phys. Rev. A {\bf 67}, 042316 (2003).
\bibitem{ern} R. R. Ernst, G.Bodenhausen, And A. Wokaun, Principles of Nuclear Magnetic Resonance in One and Two Dimensions, Oxford 
University Press 1987. 
\end{thebibliography}
\end{document}